\documentstyle[twoside,fleqn,espcrc2,epsf]{article} 
 
\newcommand{\AmS}{{\protect\the\textfont2 
  A\kern-.1667em\lower.5ex\hbox{M}\kern-.125emS}} 
 
\def\Preprint{\vspace*{-8.5cm}   
 \noindent FTUV/99-63 \\ 
  IFIC/99-66 \\   
 \vspace{6.25cm}
 }

\newcommand\mnpbp[3]{ 
{{\it Nucl. Phys. }{\bf B #1} {\it(Proc. Suppl.)} (#2) #3}} 
\newcommand{\be}{\begin{equation}} 
\newcommand{\ee}{\end{equation}} 
\newcommand{\beqn}{\begin{eqnarray}} 
\newcommand{\eeqn}{\end{eqnarray}} 
\newcommand{\dis}{\displaystyle} 
\newcommand{\ba}{\begin{array}{c}} 
\newcommand{\ea}{\end{array}} 
\newcommand{\cF}{{\cal F}} 
\newcommand{\no}{\nonumber} 
 
\def\npb#1#2#3{{\it Nucl. Phys.} {\bf B#1} (#2) #3} 
\def\plb#1#2#3{{\it Phys. Lett.} {\bf B#1} (#2) #3} 
\def\prd#1#2#3{{\it Phys. Rev.} {\bf D#1} (#2) #3} 
\def\prl#1#2#3{{\it Phys. Rev. Lett.} {\bf #1} (#2) #3} 
\def\jhep#1#2#3{{\it JHEP} {\bf #1} (#2) #3}

\def\hepph#1{hep-ph/#1} 
 
\def\hepex#1{hep-ex/#1}

\title{Tau--Decay Determination of the Strange Quark Mass 
} 

\author{Antonio Pich\address{Departament de F\'{\i}sica Te\`orica, 
 IFIC, Universitat de Val\`encia -- CSIC,\\
Dr. Moliner 50,  E-46100 Burjassot (Val\`encia), Spain\protect\thanks{
Invited talk at QCD'99, Montpellier, July 1999}
}
and Joaquim Prades\address{Departamento de  
F\'{\i}sica Te\'orica y del Cosmos, Universidad de Granada,\\ 
Campus de Fuente Nueva, E-18002 Granada, Spain}}
 
\begin{document} 
\begin{abstract} 
\noindent 
The recent ALEPH measurements of the inclusive Cabibbo--suppressed 
decay width of the $\tau$ and several moments of its invariant mass 
distribution are used to determine the value of the strange quark mass. 
We obtain,  in the $\overline{\rm MS}$ scheme, 
$m_s(M_\tau^2) = (119\pm 24)$ MeV, 
which corresponds to  
$ m_s(1\, {\rm GeV}^2) = (164 \pm 33) \,\,  {\rm MeV} ,    
\, m_s(4\, {\rm GeV}^2) = (114\pm 23) \,\, {\rm MeV} $. 
\end{abstract} 
 
\maketitle 
\Preprint
 
\section{INTRODUCTION}

The precise numerical value of the strange quark mass is a 
controversial issue, with important implications 
for low--energy phenomenology. 
The Particle Data Group \cite{PDG98} quotes a rather wide range of 
$m_s$ values, reflecting the large uncertainties in the present 
determinations of this parameter from QCD Sum Rules and Lattice 
calculations. 
 
The high precision data on tau decays \cite{TAU98} 
collected at LEP and CESR 
provide a very powerful tool to analyse strange quark mass effects 
in a cleaner environment. 
The QCD analysis of the inclusive tau decay width, 
\be 
\label{defrtau} 
R_\tau \equiv \frac{\dis \Gamma \left[ 
\tau^-\to\nu_\tau + {\rm hadrons} \;  (\gamma) \right]} 
{\dis \Gamma \left[ 
\tau^- \to e^- \, \overline{\nu}_e \, \nu_\tau (\gamma)  
\right]} \, ,  
\ee 
has already made possible \cite{PIC97} an accurate  
measurement of the strong coupling constant at the $\tau$ mass scale, 
$\alpha_s(M_\tau^2)$, 
which complements and competes in accuracy with the high precision 
measurements of $\alpha_s(M_Z^2)$ performed at LEP. 
More recently, detailed experimental studies of the Cabibbo--suppressed 
width of the $\tau$ have started to become available \cite{ALEPH99,CDH99}, 
allowing to perform
a systematic investigation of the corrections 
induced by the strange quark mass in the $\tau$ decay width 
\cite{PP98,PP99,CKP98}.  
 
What makes a $m_s$ determination from $\tau$ data very interesting is that 
the hadronic input does not depend on any extra hypothesis; 
it is a purely experimental issue, which accuracy can be 
systematically improved. The major part of the 
uncertainty will eventually come from the theoretical side. However, 
owing to its inclusive character, the total 
Cabibbo--suppressed tau decay width can be rigorously analyzed 
within QCD, using the Operator Product Expansion (OPE). 
Therefore, the theoretical input is in principle under control 
and the associated uncertainties can be quantified.

\section{THEORETICAL FRAMEWORK} 
 
The theoretical analysis of the inclusive hadronic tau decay width 
\cite{BNP92,NP88,BRA89,DP92} involves the two--point correlation functions  
$$ 
\Pi^{\mu\nu}_{ij,{\cal J}}(q) \equiv  i {\dis \int }{\rm d}^4 x\,  
e^{iqx} \, \langle 0 | T \left( {\cal J}_{ij}^\mu (x)\,  
{\cal J}_{ij}^\nu (0)^\dagger \right) |0 \rangle   
$$
%
for the vector, 
${\cal J}_{ij}^\mu = V_{ij}^\mu (x) \equiv   
\overline q_j \gamma^\mu q_i $,   
and axial--vector, 
${\cal J}_{ij}^\mu = A_{ij}^\mu (x) \equiv  \overline q_j 
\gamma^\mu\gamma_5 q_i $,  
colour--singlet quark currents 
($i,j = u, d, s$). 
These correlators have the Lorentz decompositions 
\beqn 
\Pi^{\mu\nu}_{ij,V/A}(q) &\!\! = &\!\! 
\left( - g^{\mu\nu}\, q^2 + q^\mu q^\nu \right) 
\, \Pi^{T}_{ij,V/A}(q^2)  
\no\\ &&\!\! \mbox{} + q^\mu q^\nu \, 
\Pi^{L}_{ij,V/A}(q^2) \, , 
\eeqn 
where the superscript in the transverse and longitudinal components 
denotes the corresponding angular momentum $J=1$ (T) and $J=0$ (L) 
in the hadronic rest frame.

\begin{table*} {\centering 
\begin{tabular}{|c|c|c|} 
\hline 
$(k,l)$ & $\cF^{kl}_{L+T}(x)$ & $\cF^{kl}_{L}(x)$ 
\\ \hline  (0,0) & $(1-x)^3\, (1+x)$ &  $(1-x)^3$  \\ (1,0) & $\frac{1}{10}\, 
(1-x)^4\, (7+8x)$ &  $\frac{3}{4}\, (1-x)^4$ \\ (2,0) & $\frac{2}{15}\, 
(1-x)^5\, (4+5x)$ &  $\frac{3}{5}\, (1-x)^5$ \\ (1,1) & $\frac{1}{6}\, (1-x)^4\, 
(1+2x)^2$ &   
    $\frac{3}{20}\, (1-x)^4\, (1+4x)$ \\ (1,2) & $\frac{1}{210}\, (1-x)^4\, 
(13+52 x + 130 x^2 + 120 x^3)$ &   
    $\frac{1}{20}\, (1-x)^4\, (1+4 x + 10 x^2)$ \\ \hline 
\end{tabular} 
\caption{Explicit values of the relevant kinematical kernels.} 
\label{tab:kernels}} 
\end{table*} 
 
The semi-hadronic decay rate of the $\tau$ lepton, 
can be expressed as an integral of the spectral functions 
 ${\rm Im} \, \Pi^T(s)$ and ${\rm Im} \, \Pi^L(s)$ over the invariant  
mass $s$ of the  final--state hadrons as follows: 
\beqn 
\label{rtau} 
\lefteqn{R_\tau = 12 \pi {\dis \int^{M_\tau^2}_0} \frac{{\rm d} s}{M_\tau^2} 
\, \left(1-{s\over M_\tau^2}\right)^2}&& \no \\ 
&& \mbox{}\times \left[  
\left( 1+2{s\over M_\tau^2}\right) {\rm Im}\, \Pi^T(s) 
+ {\rm Im} \, \Pi^L(s) \right] .  
\eeqn 
Moreover, according to the quantum numbers content of the 
two--point function correlators    
\beqn 
\label{correlators} 
\Pi^J(s) &\!\!\equiv&\!\! 
 |V_{ud}|^2 \left[ \Pi_{V, ud}^J(s) + \Pi_{A,ud}^J(s)\right] 
\nonumber \\  
&\!\! +&\!\! |V_{us}|^2 \left[ \Pi_{V, us}^J(s) + \Pi_{A,us}^J(s)\right] , 
\eeqn 
we can decompose $R_\tau$ into 
\beqn 
R_\tau \equiv R_{\tau, V} + R_{\tau, A} + R_{\tau, S} \, , 
\eeqn 
where $R_{\tau, V}$ and $R_{\tau, A}$ correspond to the first two terms in 
Eq.~(\ref{correlators}), while $R_{\tau, S}$ contains the remaining 
Cabibbo--suppressed contributions. 
 
The measurement of the 
invariant mass distribution of the final hadrons  
provides additional information on the QCD dynamics, through the moments 
\cite{DP92} 
\be\label{eq:momdef} 
R_\tau^{kl} \equiv \int_0^{M_\tau^2} \, ds \,  
\left( 1 -\frac{s}{M_\tau^2} \right)^k\,  
\left(\frac{s}{M_\tau^2}\right)^l \, 
{d R_\tau\over d s} \, ,
\ee 
which include  
$R_\tau\equiv R_\tau^{00}$ as a particular case. 
 
Exploiting the analytic properties of $\Pi^J(s)$, 
we can express these moments  
as  contour integrals in the complex 
$s$-plane running counter-clockwise around the circle $|s|=M_\tau^2$: 
\beqn 
\label{contourkl} 
R_\tau^{kl} &\!\!\!\! =&\!\!\!\! -\pi i \oint_{|x|=1}  \frac{{\rm d}x}{x} 
\,\Bigl\{ 3 \, \cF^{kl}_{L+T}(x) \, D^{L+T}(M_{\tau}^2 x) 
\Bigr.\no\\ &&\qquad\qquad\quad\; \Bigl.\mbox{} 
+ 4 \, \cF^{kl}_L(x) \, D^L(M_{\tau}^2 x) \Bigr\}  . \; 
\eeqn 
We have used integration by parts to rewrite $R_\tau^{kl}$ in terms of  
the logarithmic derivatives 
%
\be 
D^{L+T}(s)\,\equiv\,  -s \frac{{\rm d}}{{\rm d}s} 
\left[\Pi^{L+T}(s)\right] 
\, ,  
\ee\be 
D^{L}(s) \,\equiv\, \frac{\dis s}{\dis M_\tau^2} \, 
\frac{\dis {\rm d}}{\dis {\rm d}s}  \left[s\, \Pi^{L}(s)\right] \, , 
\ee 
which satisfy homogeneous renormalization group equations. 
All kinematical factors have been absorbed into the kernels 
$\cF^{kl}_{L+T}(x)$ and $\cF^{kl}_L(x)$. 
Table~\ref{tab:kernels} shows the explicit form of these kernels for 
the moments which we are going to analyze in the following sections.

For large enough $-s$, the contributions to $D^J(s)$ 
can be organized with the OPE  
in a series of local gauge--invariant scalar operators 
of increasing dimension $D=2n$, 
times the appropriate inverse powers of $-s$. 
This expansion is expected to be well behaved 
along the complex contour $|s|=M_\tau^2$, except in the crossing point with 
the positive real axis \cite{PQS76}. 
 As shown in Table~\ref{tab:kernels}, the region 
near the physical cut is strongly suppressed by a zero of 
order $3+k$ at $s=M_\tau^2$. Therefore, the uncertainties 
associated with the use of the OPE near the time--like axis are 
very small. 
Inserting this series in (\ref{contourkl}) and evaluating the contour  
integral, one can rewrite $R_\tau^{kl}$ as an expansion in inverse 
powers of $M_\tau^2$ \cite{BNP92}, 
\beqn 
\label{LAB:deltas} 
\lefteqn{R_\tau^{kl} \equiv  
3 \left[ |V_{ud}|^2 + |V_{us}|^2 
\right] S_{\rm EW} \,\biggl\{ 1 + \delta'_{\rm EW}\, + \delta^{kl\, (0)} 
\biggr. }\no\\ &&\!\!\!\!\!\! \biggl.\mbox{} + 
{\dis \sum_{D=2,4,\cdots}} \left( \cos^2{\theta_C} \, \delta_{ud}^{kl\, (D)}+ 
\sin^2{\theta_C} \, \delta_{us}^{kl\, (D)}  \right) \biggr\} , \no 
\eeqn 
where $\sin^2{\theta_C}\equiv |V_{us}|^2/[|V_{ud}|^2+|V_{us}|^2]$ 
and we have pulled out the electroweak 
corrections  $S_{\rm EW}=1.0194$   \cite{MS88}  
and $\delta'_{\rm EW}\simeq 0.0010$ 
\cite{BL90}.  
 
The dimension--zero contribution $\delta^{kl\, (0)}$  
is the purely perturbative correction, neglecting quark masses, 
which, owing to chiral symmetry, is identical for the vector and  
axial--vector parts. 
The symbols $\delta_{ij}^{kl\, (D)} \equiv [ \delta^{kl\, (D)}_{ij,V} 
+ \delta^{kl\, (D)}_{ij,A}]/2$ stand for the average of the vector and  
axial--vector contributions from dimension $D\ge 2$ operators; they 
contain an implicit suppression factor $1/M_\tau^D$.

\section{SU(3) BREAKING} 
 
The separate measurement of the Cabibbo--allowed and Cabibbo--suppressed 
decay widths of the $\tau$ \cite{ALEPH99} allows one to pin down the 
SU(3) breaking effect induced by the strange quark mass, through 
the differences 
\beqn 
\delta R_\tau^{kl} &\!\!\equiv &\!\! {R_{\tau,V+A}^{kl}\over |V_{ud}|^2} - 
{R_{\tau,S}^{kl}\over |V_{us}|^2}  
\no\\ &\!\! =&\!\! 3 \, S_{EW}\,\sum_{D\geq 2}  
\left[ \delta^{kl\, (D)}_{ud} - \delta^{kl\, (D)}_{us} \right] 
\, . 
\eeqn 
%
 
The leading contributions to $\delta R_\tau^{kl}$ are 
quark--mass corrections of dimension two 
\cite{PP98,PP99}; they 
are the dominant SU(3) breaking effect, generating the wanted 
sensitivity to the strange quark mass. 
The corrections of $O(m^4)$ are very tiny \cite{PP99}. 
The main $D=4$ contribution comes from the SU(3)--breaking 
quark condensate 
\be\label{eq:O4def} 
\delta O_4 \,\equiv\, \langle 0| m_s\, \bar s s 
  - m_d \,\bar d d | 0 \rangle \, . 
\ee 
Neglecting the small $O(m^4)$ terms and $D\ge 6$ contributions, 
$\delta R_\tau^{kl}$ can be written as \cite{PP99}: 
\beqn 
\label{Delta2a} 
\delta R_\tau^{kl} &\!\!\approx &\!\! 24\, S_{EW}\,\biggl\{ 
{m_s^2(M_\tau^2)\over M_\tau^2} \, \left(1-\epsilon_d^2\right)\, 
\Delta^{(2)}_{kl}(a_\tau)  
\biggr.\no\\ &&\biggl.\qquad\qquad 
- 2 \pi^2\,  {\delta O_4\over M_\tau^4} \, Q_{kl}(a_\tau) 
\biggr\} \, ,
\eeqn 
where 
$\epsilon_d\equiv  m_d/ m_s = 0.053 \pm 0.002 $ \cite{LEU96} and  
$a_\tau\equiv \alpha_s(M_\tau^2)/\pi$.

\begin{table}[thb] 
\centering 
\begin{tabular}{|c|c|c|} 
\hline  
$(k,l)$ & $\Delta^{(2)}_{kl}(a_\tau)$  & $Q_{kl}(a_\tau)$ 
\\ \hline 
(0,0) & $2.0 \pm 0.5$  & $1.08\pm 0.03$ \\ 
(1,0) & $2.4 \pm 0.7$  & $1.52\pm 0.03$ \\ 
(2,0) & $2.7 \pm 1.0$  & $1.93\pm 0.02$ \\ 
(1,1) & $-0.39\pm 0.26$  & $-0.41\pm 0.02$ \\ 
(1,2) & $0.07\pm 0.06$  & $-0.02\pm 0.01$ 
\\ \hline 
\end{tabular} 
\caption{Numerical values \protect\cite{PP99} 
of the relevant perturbative expansions for 
$\alpha_s(M_\tau^2) = 0.35\pm0.02$.  
}  
\label{tab:num} 
\end{table} 
 
The perturbative QCD expansions $\Delta^{(2)}_{kl}(a_\tau)$ 
and $Q_{kl}(a_\tau)$ are known to $O(a_\tau^2)$. 
Moreover, the $O(a_\tau^3)$ contributions to $\Delta^{(2)}_{kl}(a_\tau)$ 
coming from the longitudinal correlator $D^L(s)$ 
have been also computed. 
Using the value of the ($\overline{\mbox{\rm MS}}$) 
strong coupling determined by the total hadronic $\tau$ decay 
width \cite{PIC97}, 
$\alpha_s(M_\tau^2) =  0.35 \pm0.02$, 
one gets the numerical results shown in Table~\ref{tab:num} \cite{PP99}. 
 
The rather large theoretical uncertainties of 
\be 
\Delta^{(2)}_{kl}(a_\tau)\equiv  
 {1\over 4} \,\left\{ 3 \,\Delta^{L+T}_{kl}(a_\tau) 
+ \Delta^{L}_{kl}(a_\tau) \right\} \, ,  
\ee
have their origin in the bad 
perturbative behaviour of the longitudinal contribution. 
The most important higher--order corrections can be resummed \cite{PP98}, 
using the renormalization group,  
but the resulting ``improved'' series is still rather badly behaved. 
For instance, 
$$ 
\Delta^{L}_{00}(0.1) = 1.5891 + 1.1733 + 1.1214 + 1.2489 + \cdots 
$$ 
which has $O(a^2)$ and $O(a^3)$ contributions of the same size. 
On the contrary, the $J=L+T$ series converges very well: 
$$ 
\Delta^{L+T}_{00}(0.1) = 0.7824 + 0.2239 + 0.0831  
+ \cdots 
$$ 

Fortunately, the longitudinal contribution to $\Delta^{(2)}_{kl}(a_\tau)$ 
is parametrically suppressed by a factor $1/3$. Thus, the combined final 
expansion looks still acceptable for the first few terms: 
\beqn 
\lefteqn{\Delta^{(2)}_{00}(0.1) = 0.9840 + 0.4613 + 0.3427}&& 
\no\\ &&\mbox{}
 + \left( 0.3122 - 0.000045\, c_3^{L+T}\right) + \cdots 
\eeqn 
Nevertheless, after the third term the series appears to be dominated 
by the longitudinal contribution, and the bad perturbative behaviour 
becomes again manifest. 
Taking the unknown $O(a^3)$ coefficient of the 
$D^{L+T}(s)$ perturbative series as 
$c_3^{L+T} \sim c_2^{L+T}\,\left(c_2^{L+T}/c_1^{L+T}\right)\approx 
323$, 
the fourth term becomes $0.298$; i.e. a 5\% reduction only. 
 
Since the longitudinal series  seems to 
reach an asymptotic behaviour at $O(a^3)$, 
the central values of $\Delta^{(2)}_{kl}(a_\tau)$ 
have been evaluated   
adding to the fully known $O(a^2)$ result one half of the  
longitudinal $O(a^3)$ contribution. 
To estimate the associated  
theoretical uncertainties, we have taken one half of the size of the 
last known perturbative contribution plus the variation induced 
by a change of the renormalization scale in the range 
$\xi\in [0.75,2]$ (added in quadrature). 
 
 
The SU(3)--breaking condensate $\delta O_4$ could be 
extracted from the $\tau$ decay data, together with $m_s$, through 
a combined fit of different $\delta R_\tau^{kl}$ moments. 
However, this is not possible with the actual  
experimental accuracy.  
We can estimate the value of  $\delta O_4$  
using the constraints provided by chiral symmetry.   
To lowest order in Chiral Perturbation Theory, 
 $\delta O_4$ 
is fully predicted in terms of the pion decay  
constant and the pion and kaon masses: 
$\delta O_4 \simeq 
-f_\pi^2\, \left(m_K^2 - m_\pi^2\right) \simeq -1.9\times 10^{-3} 
\:\mbox{\rm GeV}^4  
$.  
%
Taking into account the leading 
$O(p^4)$ corrections through the ratio of quark vacuum  
condensates \cite{NAR89,DJN89}   
\be 
v_s \,\equiv\, \frac{\langle 0 | \overline s s | 0 \rangle} 
{\langle 0 | \overline d d | 0\rangle}\, =\, 0.8 \pm 0.2 \, , 
\ee 
%
one gets the improved estimate, 
\beqn\label{eq:O4value} 
\delta O_4  &\!\! \simeq &\!\! 
- \frac{m_s}{2 \hat m}\, (v_s -\epsilon_d)\, 
\, f_\pi^2 \, m_\pi^2  
\nonumber \\  
&\!\!\simeq &  \!\! 
 -(1.5\pm 0.4)\times 10^{-3} 
\:\mbox{\rm GeV}^4 \, ,  
\eeqn 
where we have used the known quark mass ratio 
\cite{LEU96} 
$m_s/\hat m = 24.4\pm 1.5$.    
 
Strictly speaking, $\delta O_4$ and $v_s$ are scale dependent. 
This dependence cancels with the $O(m^4)$ contributions \cite{PP99} 
and is then of $O(p^8)$ in the chiral expansion. The numerical  
effect is smaller than the accuracy of (\ref{eq:O4value}) and has 
been neglected together with the tiny $O(m^4)$ corrections.

\section{NUMERICAL ANALYSIS} 
\label{sec:numerics} 
 
\begin{table}[t]  
\centering 
\begin{tabular}{|c|c|c|} 
\hline 
$(k,l)$ & $\delta R_\tau^{kl}$ & $m_s(M_\tau^2)$ (MeV) 
\\ \hline  
(0,0) & $0.394\pm 0.137$ & $143\pm31\pm18$\\ 
(1,0) & $0.383\pm 0.078$ & $121\pm17\pm18$\\ 
(2,0) & $0.373\pm 0.054$ & $106\pm12\pm21$\\ 
(1,1) & $0.010\pm 0.029$ & --\\ 
(1,2) & $0.006\pm 0.015$ & --\\ \hline 
\end{tabular} 
\caption{Measured \protect{\cite{ALEPH99}} moments $\delta R_\tau^{kl}$ 
and corresponding $m_s(M_\tau^2)$ values \protect{\cite{PP99}}. 
The first error is experimental 
and the second theoretical.}
\label{tab:res} 
\end{table} 
 
The ALEPH collaboration has measured \cite{ALEPH99} the weighted  
differences $\delta R_\tau^{kl}$ for five different values of  
$(k,l)$. The experimental results are shown in Table~\ref{tab:res}, 
together with the corresponding $m_s(M_\tau^2)$ values. 
Since the QCD counterparts to the moments $(k,l)=$ (1,1) and (1,2) 
have theoretical uncertainties larger than 100\%,  
we only use  the  moments 
$(k,l)=$ (0,0), (1,0), and (2,0).

The experimental errors quoted in Table~\ref{tab:res} do not include 
the present uncertainty in $|V_{us}|$. 
To estimate the corresponding error in $m_s$, we take the following 
numbers from ALEPH: 
$R^{00}_{\tau, V+A}=3.486\pm0.015$,  
$R^{00}_{\tau, S}=0.1610\pm0.0066$, 
$|V_{ud}|=0.9751\pm0.0004$ and $|V_{us}|=0.2218\pm0.0016$. 
This gives 
$\delta R_\tau^{00}=0.394\pm0.135\pm0.047$, 
where the second error comes from the uncertainty in $|V_{us}|$ 
and translates into an additional uncertainty of 10 MeV  
in the strange quark mass. 
We will put the same $|V_{us}|$ 
uncertainty to the other two moments, 
for which the ALEPH collaboration does not quote the separate values of 
$R^{kl}_{\tau, V+A}$ and $R^{kl}_{\tau, S}$.

Taking the information from the three moments into account, 
we get our final result \cite{PP99}: 
\beqn\label{eq:result} 
m_s(M_\tau^2) &\!\! = &\!\! (119 \pm 12 \pm 18 \pm 10) \: {\rm MeV}  
\no\\ &\!\! = &\!\! (119 \pm 24 ) \: {\rm MeV} \, . 
\eeqn 
The first error is experimental, the second reflects the QCD 
uncertainty and the third one 
is from the present uncertainty in $|V_{us}|$. 
Since the three moments are highly correlated, we have taken the 
smaller individual errors as errors of the final average. 
%
Our determination (\ref{eq:result}) corresponds to 
\be 
m_s(1\, {\rm GeV}^2)  
= (164 \pm 33 ) \:{\rm MeV}  
\ee 
and  
\be 
m_s(4\, {\rm GeV}^2)  
= (114 \pm 23 ) \: {\rm MeV} \, . 
\ee 
 
\section{COMPARISON WITH ALEPH} 
 
The ALEPH collaboration has performed a phenomenological 
analysis of the $\delta R_\tau^{kl}$ moments in Table~\ref{tab:res}, 
which results in larger $m_s$ values \cite{ALEPH99}: 
$$ 
m_s(M_\tau^2)  = \left\{ 
\begin{array}{l} 
 149 {}^{+24_{\rm exp}}_{-30_{\rm exp}} 
 {}^{+21_{\rm th}}_{-25_{\rm th}} \pm 6_{\rm fit} \: {\rm MeV}  ,
\\ 
  176 {}^{+37_{\rm exp}}_{-48_{\rm exp}} 
 {}^{+24_{\rm th}}_{-28_{\rm th}} \pm 8_{\rm fit} 
\pm 11_{J=0}  \: {\rm MeV}  .
\end{array}  \right. 
$$ 
To derive these numbers, ALEPH has used our published results in 
refs.~\cite{PP98}, \cite{BNP92} and \cite{DP92}. 
Since we have analyzed the same data with improved theoretical 
input \cite{PP99}, it is worthwhile to understand the 
origin of the numerical difference. 
 
ALEPH makes a global fit to the five measured moments, including 
the last two which are unreliable (100\% theoretical errors). 
In view of the asymptotic behaviour of $\Delta_{kl}^L(a_\tau)$, 
they truncate this perturbative series at $O(a_\tau)$, 
neglecting the known and positive 
$O(a_\tau^2)$ and $O(a_\tau^3)$ contributions. Thus, they use a 
smaller value of $\Delta_{kl}^{(2)}(a_\tau)$ and, therefore, get 
a larger result for $m_s$  
(the first value above) 
because the sensitivity to this parameter 
is through the product $m_s^2(M_\tau^2) \Delta_{kl}^{(2)}(a_\tau)$. 
Since they put rather conservative errors, their result is 
nevertheless consistent with ours. 
 
ALEPH has made a second analysis 
subtracting the $J=L$ contribution. Unfortunately, only the 
pion and kaon contributions are known.  
Using the positivity of the longitudinal spectral functions, 
this pole contributions provide lower bounds on 
$\mbox{\rm Im}\Pi^L_{ud}(s)$ and $\mbox{\rm Im}\Pi^L_{us}(s)$, 
which translate into lower limits on the corresponding $J=L$ 
contribution to $\delta R_\tau^{kl}$.
Subtracting this contribution, one gets upper bounds on 
$\delta R_{\tau,L+T}^{kl}$ \cite{PP99} which imply  
$m_s(M_\tau^2) < 202$ MeV \cite{PP99}. 
 
However, besides subtracting the pion and kaon poles, ALEPH 
makes a tiny ad-hoc correction to account for the remaining 
unknown $J=L$ contribution, and quotes the resulting number 
as a $m_s(M_\tau^2)$ determination [the second value above]. 
Since they add a generous uncertainty, their number does not 
disagree with ours. However, it is actually an 
upper bound on $m_s(M_\tau^2)$ and not a determination of this 
parameter.

\section*{ACKNOWLEDGEMENTS}

We have benefit from useful discussions with
Shaomin Chen, Michel Davier, Andreas H\"ocker and Edwige Tournefier. 
This work has been supported in part by the
European Union TMR Network EURODAPHNE (Contract No. ERBFMX-CT98-0169),
by CICYT, Spain, under grants   No. AEN-96/1672
and PB97-1261, and by 
Junta de Andaluc\'{\i}a, Grant No. FQM-101.

\vfill

\end{document}